\documentclass[aps,prl,twocolumn,groupedaddress,showpacs]{revtex4}
\usepackage{graphicx}
\usepackage{amssymb}
\usepackage{color}

\begin{document}



\title{Three-body recombination in a three-state Fermi gas with widely tunable interactions}


\author{J. H. Huckans, J. R. Williams, E. L. Hazlett, R. W. Stites, and K. M. O'Hara}
\affiliation{Department of Physics, Pennsylvania State University,
University Park,\nolinebreak \,Pennsylvania 16802-6300, USA}


\date{\today}

\begin{abstract}
We investigate the stability of a three spin state mixture of ultracold fermionic $^6$Li
atoms over a range of magnetic fields encompassing three Feshbach resonances.
For most field values,
we attribute decay of the atomic population to three-body processes involving one atom from each spin
state and find that the three-body loss coefficient varies by over four orders of magnitude.
We observe
high stability when at least two of the three scattering lengths are small,
rapid loss near the Feshbach resonances, and two unexpected resonant loss features.
At our highest fields, where all pairwise scattering lengths are approaching
$a_t = -2140 a_0$, we measure a three-body loss coefficient
$L_3 \simeq 5\times 10^{-22}\,\mathrm{cm}^6/\mathrm{s}$ and a trend toward lower decay
rates for higher fields indicating that future studies of color superfluidity and trion
formation in a SU(3) symmetric Fermi gas may be feasible.
\end{abstract}


\maketitle

%
%
Multi-component Fermi gases with tunable interactions are exceptionally well suited to
the study of few- and many-body quantum physics. Ultracold two-state Fermi gases near
a Feshbach resonance have been used to characterize the crossover from
Bardeen-Cooper-Schreiffer (BCS)
superfluidity to Bose-Einstein condensation (BEC) of diatomic
molecules~\cite{BCSBECList}.
Further, a normal to superfluid transition and phase separation have been observed in
imbalanced
two-state spin mixtures~\cite{ImbalanceList}.
The stability of the two-state mixtures against two- and three-body loss processes
was critically important to the success of these experiments.

%
Recently, there has been considerable interest in the study of three-state Fermi gases
with tunable
interactions~\cite{SU3List, Zhuang06, Hofstetter07, Efimov, AssortThreeFermi, White08}.
Whereas three-body interactions in ultracold two-state Fermi gases are suppressed by
the exclusion principle, the addition of a third spin component allows for the study
of three-body phenomena such as the Efimov effect~\cite{Efimov}.  Further,
with this additional spin state there may be competition between multiple pairing
states and trion formation~\cite{SU3List, Zhuang06, AssortThreeFermi}.
For example, if pairwise interactions are all attractive and of equal magnitude, the
system is expected to exhibit a novel superfluid phase analogous to color superconductivity
 in quantum chromodynamics (QCD)~\cite{Hofstetter07}.  A quantum phase transition to a Fermi liquid of trimers
(analogous to baryons) may be observed if the ratio of the
interaction energy to the kinetic energy is increased (e.g. by an optical lattice)~\cite{Hofstetter07}.

%
%

Future studies of the above phenomena critically depend on the magnitude of
two- and three-body loss rates,
particularly when two or more scattering lengths are resonantly enhanced.
Two- and three-body loss and heating processes can impose stringent limits on the maximum achievable
phase space density.  Further, the unambiguous observation of three-body resonances
requires negligible two-body loss~\cite{Grimm06}.

Most investigations of Fermi gases with tunable interactions have focussed on
two-state mixtures.
A small admixture of a third spin component has been used for
thermometry~\cite{Regal05} and
recently the rapid decay of a three-state Fermi gas near a Feshbach
resonance was noted during an
investigation of Cooper pair size by radio-frequency (RF) spectroscopy
in a two-state Fermi gas~\cite{Ketterle08}.  Very
recently our group reported measurements of three-body loss in thermal
and degenerate three-state
Fermi gases of $^6$Li for magnetic fields between 400 and 960~G which included three Feshbach
resonances~\cite{Huckans08}.  Independently, Ottenstein {\it et al.}
observed loss in a three-state $^6$Li gas for field values between 0 - 750~G (including a single
Feshbach resonance) and reported three-body loss rate coefficients for field values up to 600~G (excluding any pairwise scattering resonances)~\cite{Jochim08}.


%
%
In this Letter, we report the first measurements of three-body loss coefficients
for a three-state
Fermi gas over a range of magnetic fields where pairwise interactions are
resonantly enhanced.  This enhancement
is due to three overlapping Feshbach resonances and a zero-energy resonance in the $^6$Li
triplet molecular potential.  Further, we demonstrate that all
two-state mixtures of the three
lowest energy hyperfine states of $^6$Li are stable against two-body
loss processes for almost the entire range of field
values between 15 and 953~G, making this three-state mixture well suited to studies of three-body physics.
We observe a narrow loss feature at 127~G~\cite{JochimObservation} which may be due
to a three-body resonance and an additional state-dependent loss feature at 504~G.  Finally,
we measure the three-body loss rate coefficient at high fields
where all three scattering lengths are asymptoting to the
triplet scattering length $a_t = -2140 a_0$.  This measurement
and an observed trend toward lower loss at higher
fields has important implications for realizing cold atom analogs
of color superfluid and baryon phases in QCD.

\begin{figure}
 \includegraphics[width=3.2in]{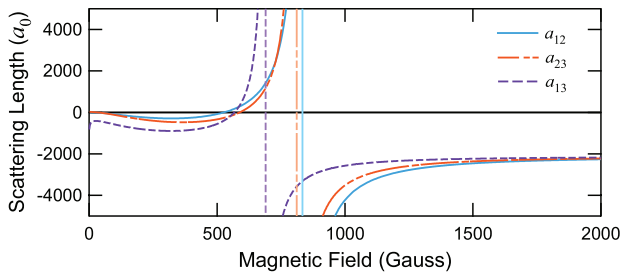}
 \caption{\label{ThreeStateProps}  (color online) The $s$-wave scattering lengths as a function of magnetic
 field for interactions between the three lowest energy hyperfine ground states of $^6$Li ($|1\rangle$, $|2\rangle$ and $|3\rangle$)~\cite{Julienne05}.
 Feshbach resonances occur at 690, 811, and 834~G.\vspace{-0.15in}}
\end{figure}

%
%

We study a $^6$Li Fermi gas with equal populations in the three lowest energy
hyperfine states.  At zero field, the three states correspond to
$|1\rangle = |\frac{1}{2}, +\frac{1}{2}\rangle$,
$|2\rangle = |\frac{1}{2},-\frac{1}{2}\rangle$ and
$|3\rangle = |\frac{3}{2},-\frac{3}{2}\rangle$
in the $|f,m_f\rangle$ basis.  For fields above $\simeq 200\,\mathrm{G}$, these states
become increasingly electron-spin polarized and primarily differ by their nuclear spin projection.
The pairwise $s$-wave scattering lengths between these states
($a_{12}$, $a_{23}$, and $a_{13}$)
exhibit three broad overlapping Feshbach resonances as shown in
Fig.~\ref{ThreeStateProps}~\cite{Julienne05}.
For very large magnetic fields, each of the pairwise scattering
lengths asymptote to the
triplet scattering length for $^6$Li, $a_t = -2140 a_0$.  Therefore,
in the limit of large
magnetic fields, the interactions are SU(3) symmetric~\cite{Hofstetter07}.

%
%
Two-body spin-flip processes are expected to be especially small for this ultracold
$^6$Li mixture.  Spin-exchange collisions are energetically forbidden in a magnetic field.
Dipolar relaxation by electron spin-flip is suppressed at high field as the gas
becomes increasingly electron-spin polarized in the lowest energy electron spin state.

%
%
We can produce a degenerate Fermi gas (DFG) with an equal mixture of atoms in states
$|1\rangle$ and $|2\rangle$
once every 5 seconds by evaporatively cooling this mixture in a red-detuned optical dipole trap~\cite{Thomas03}.
During the first second, $\sim 10^8$ $^6$Li atoms from a Zeeman-slowed atomic
beam are collected in a magneto-optical trap (MOT). A crossed optical dipole trap which overlaps the MOT is then
turned on and optimally loaded by tuning the MOT laser beams $\simeq 6$ MHz below resonance and reducing their
intensity for 7 ms.  The atoms are then optically pumped into states $|1\rangle$ and $|2\rangle$ prior to
extinguishing the MOT laser beams and field gradient.  The two beams which form the optical trap derive from a
linearly polarized multi-longitudinal-mode 110 W fiber laser operating at $1064\,\mathrm{nm}$.  The beams nearly
co-propagate in the vertical direction ($\hat{y}$), have orthogonal polarizations and cross at an angle $\simeq 11^\circ$.
The beams are elliptical with calculated $e^{-2}$ waist radii $\sim 30 \mu{\mathrm{m}}$
and $\sim 100 \mu{\mathrm{m}}$ at the point of intersection.  The maximum trap depth per beam is
$\simeq 1\,{\mathrm{mK}}$ allowing $\sim 5\times 10^6$ atoms to be initially loaded.
We apply a bias field
and a noisy RF pulse to create a 50-50 mixture of atoms in
states $|1\rangle$ and $|2\rangle$~\cite{Thomas03}.

\begin{figure}
 \includegraphics[width=3.4in]{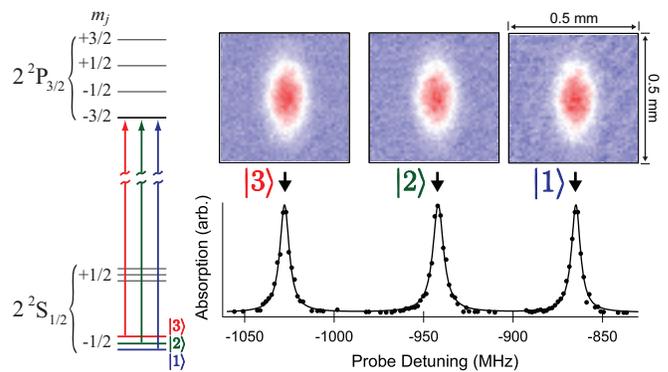}
 \caption{\label{ThreeStates} (color online) The population in states $|1\rangle$, $|2\rangle$ and $|3\rangle$
 measured by absorption imaging in the Paschen-Back regime at a bias field of 568~G~\cite{FieldUncertainty}.
 A $\sigma^{-}$ polarized probe beam drives
 $m_j = -\frac{1}{2} \rightarrow m_{j'} = - \frac{3}{2}$ transitions which, due to the hyperfine
 interaction in the ground state,  are spectroscopically resolvable.  (Top right) The on-resonance absorption images show
 the density profile for the three states following a 900~$\mu$s time-of-flight expansion.  (Bottom right) The total absorption
 vs. probe detuning (relative to the
 $|f=\frac{1}{2},m_f=-\frac{1}{2}\rangle$ $\rightarrow$ $|f'=\frac{3}{2},m_{f'}=-\frac{3}{2}\rangle$ transition frequency in zero field).\vspace{-0.15in}}
\end{figure}

Forced evaporation of the atoms occurs at a field of 330~G where the two-state
scattering length
$a_{12} \simeq -280 a_0$.  We lower the depth of the optical trap $U_0$ by a
factor of $107$ over 3.6 seconds
to obtain our final temperature.  For the work presented here we study a gas at a temperature
$T \gtrsim 0.5\,T_F$ ($T_F$ is the Fermi temperature) so that it is appropriate to treat the
cloud as a thermal gas in the analysis.  To suppress further loss by evaporation for the remainder of
the experiment, $U_0$ is increased by a factor of 4 increasing the ratio
of $U_0/k_B T$ by a factor $\simeq 2$.
The final oscillation frequencies of the
trap are $\nu_x = 3.84\,\mathrm{kHz}$, $\nu_y = 106\,\mathrm{Hz}$ and
$\nu_z = 965\,\mathrm{Hz}$~\cite{TrapFrequencies} with a final trap depth per beam
$\simeq 40\,\mu{\mathrm{K}}$.  The total number of atoms
$\simeq 3.6\times 10^5$ in
a balanced two-state mixture at $T \simeq 1.9\,\mu{\mathrm{K}}$
for which $T_F \simeq 3.7 \mu{\mathrm{K}}$.  The background limited $1/e$-lifetime
$\simeq 30\,{\mathrm{s}}$.

%
%

To create an incoherent three-state mixture we first increase the strength of the
magnetic field in 10~ms
to 568~G where the mixture is stable.  We then apply a noisy RF magnetic field
with two frequencies centered on the $|1\rangle - |2\rangle$ and
$|2\rangle - |3\rangle$ transitions (each with a power corresponding to an on-resonance Rabi frequency $\Omega \sim 2\pi\times18\,\mathrm{kHz}$
and both broadened to a width
of 1~MHz).  A magnetic
field gradient of $\simeq 1$~G/cm in the direction
of the bias field ($\hat{z}$) is simultaneously applied
to destroy any internal state coherence in the sample.  The noisy RF field
remains on for 50~ms to
create an equal mixture of $N\simeq 1.2\times10^5$ atoms in each state.  At this point,
$T = 1.9 \,\mu{\mathrm{K}}$ with $T_F = 3.2\,\mu{\mathrm{K}}$ and the peak density in a single spin state is
$5.5 \times 10^{12} \,{\mathrm{atoms/cm}}^{3}$.  We measure the population
and temperature of atoms in each state separately by spectroscopically-resolved
absorption imaging (Fig.~\ref{ThreeStates}).

%
%

\begin{figure}
 \includegraphics[width=3.1in]{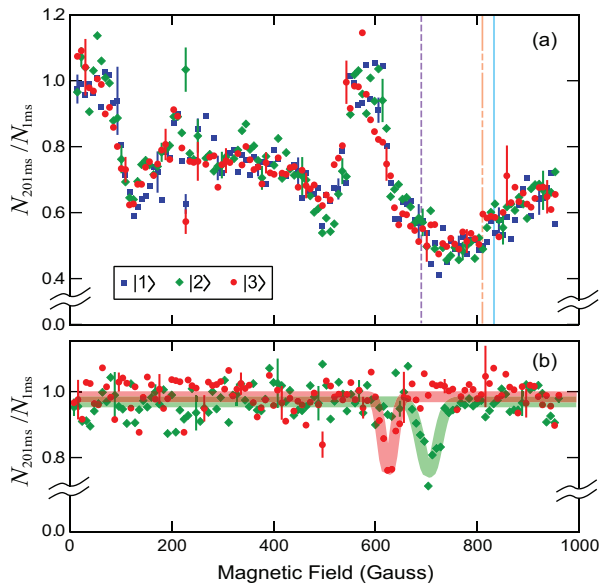}
 \caption{\label{FractionRemain} (a) Number remaining in each of the three spin states after
 spending 201~ms at the field of interest $B$ (normalized to the number remaining after spending 1~ms at $B$).
 (b) Magnetic field dependence of decay in individual two-state mixtures.  \textcolor{green}{$\blacklozenge$} (\textcolor{red}{$\bullet$})
 indicates the normalized number remaining in state $|2\rangle$ ($|3\rangle$) of a $|2\rangle-|3\rangle$ ($|1\rangle-|3\rangle$) mixture after 201~ms.
 The thick lines guide the eye.  The representative error bars indicate the standard deviation in the mean.  Absorption images are taken at
 953~G to ensure that any weakly-bound molecules which remain trapped are dissociated and measured.\vspace{-0.15in}}
\end{figure}

To investigate the magnetic field dependence of decay in this three state mixture,
we compared the number of atoms remaining in each spin state after spending two different times, 1~ms and 201~ms,
at a particular field of interest $B$ (see Fig.~\ref{FractionRemain}(a)).
For each experimental cycle the magnetic field was ramped in 10~ms from 568~G (where the mixture was first created) to
the field of interest where it was
held constant for either 1~ms or 201~ms.  The field was then ramped in 10~ms to 953~G
and held for 20~ms before absorption imaging at
953~G to measure the atom number ($N_{1{\mathrm{ms}}}$ or $N_{201{\mathrm{ms}}}$).  The 10~ms field ramp
to 953~G (where $a_{12}, a_{23}, a_{13} < 0$) ensured that any atoms which had formed weakly bound molecules
in the vibrational state associated with the Feshbach resonances (but remained trapped) would be dissociated
and measured. For each field value, $N_{201{\mathrm{ms}}}$ is normalized by $N_{1{\mathrm{ms}}}$ to correct for
the fraction of atoms lost during the field sweeps, both to and from the field of interest $B$, which exhibited a slowly varying
dependence on $B$.  See Ref.~\cite{EPAPS} for the data before normalization.

To confirm that two-body loss does not contribute significantly to the measurements described above,
we measured the ratio of atoms remaining after evolution times of 201 ms and 1 ms at
fixed magnetic field values for each of the three possible two-state mixtures.
Figure~\ref{FractionRemain}(b) shows the ratio $N_{201{\mathrm{ms}}}/N_{1{\mathrm{ms}}}$ for atoms in state $|3\rangle$
(state $|2\rangle$) of a $|1\rangle-|3\rangle$ ($|2\rangle-|3\rangle$) mixture.  Data for the $|1\rangle-|2\rangle$
mixture (not shown) is very similar to that of the $|2\rangle - |3\rangle$ mixture (e.g. Ref.~\cite{KetterleFesh}).  With the exception of
loss features between $600$ and  $750\,\mathrm{G}$, each two-state mixture was stable.
The loss features are due
to three-body recombination to bound molecular states associated with
the Feshbach resonance. Since the two-state mixtures are stable for fields $\leq 600\,\mathrm{G}$ and
$\geq 750\,\mathrm{G}$, loss features observed in three-state mixtures at these
fields are due to three-body events involving one atom from each spin state.

%
%

In Fig.~\ref{FractionRemain}(a), the broad dominant loss feature centered
at 720~G occurs in the vicinity of three overlapping interspecies Feshbach resonances.
Significant loss due
to three-body recombination is expected near these resonances since recombination
events involving one
atom in each spin state are {\emph{not}} suppressed by the exclusion principle and
one expects a significant
increase in the event rate when two or more scattering lengths are resonantly enhanced~\cite{Mediator}.
Similarly, high
stability at zero field and near the zero crossings of the Feshbach
resonances is not surprising since at least two of the scattering lengths
are small at these fields.
At high field, the stability increases relative to that at 720~G.

Unexpected resonant loss features are
observed at 127 and 504 G, where the two-body scattering lengths are not predicted
to exhibit any resonances.  We have observed these features at higher temperatures
and in a different trap configuration~\cite{EPAPS}.
A possible explanation for the feature at 127 G is that the binding energy of a
three-body bound state crosses through zero near this field value.
At 504 G, we observe differing degrees of enhanced loss for the three states
leading to population imbalance.  This resonance was not observed in Ref.~\cite{Jochim08}.

The feature at 228~G in Fig.~\ref{FractionRemain} is consistent with our estimated location of
a $|1\rangle-|3\rangle$ $p$-wave Fesh\-bach resonance  which would yield the
observed preferential loss of atoms from states $|1\rangle$ and $|3\rangle$~\cite{PWave}.
We also observed a very narrow but inconsistent loss feature at 259~G (not shown).  The modest increase in stability
observed near 200~G did not appear in our other experimental data~\cite{EPAPS}.

%
%

A second set of experiments measured three-body loss coefficients at fixed magnetic fields
where the decaying populations remained balanced.
In these experiments, the magnetic field is swept to the field of interest $B$ in 10~ms
after creation of the three-state mixture and the number $N(t)$ and
temperature $T(t)$ of atoms in state $|3\rangle$ are then measured at the field $B$ for various delay times $t$.
To extract the three-body loss coefficients, we fit this data
assuming one- and three-body loss and heating
due to three-body recombination.

%
%

When three-body recombination
involves one atom from each spin state, the number of trapped atoms in each
of the equally populated spin states $N(t)$ evolves according
to $\dot{N} = - L_1 N - L_3 \langle n^2 \rangle N$.
Here, $\langle n^2 \rangle$ is the average value of the squared density
per spin state and $L_3$ ($L_1$) is the three-body (one-body) atom-loss rate
coefficient.  The density distribution in the trap
is well described by a thermal distribution so that
\begin{eqnarray}
\label{NumEvol}
\frac{d N}{d t} & = & -L_1 N -\gamma \frac{N^3}{T^3}
\end{eqnarray}
where $\gamma = L_3 (m \bar{\omega}^2/2 \pi k_B)^3/\sqrt{27}$
and $\bar{\omega} = \left(\omega_x \omega_y \omega_z \right)^{1/3}$.
Following Ref.~\cite{Grimm03}, we model the temperature increase in
the gas as arising from ``anti-evaporation'' and recombination
heating.  The temperature evolves as
\begin{eqnarray}
\label{TempEvol}
\frac{d T}{d t} & = & \gamma \frac{N^2}{T^3}\frac{\left(T + T_h\right)}{3},
\end{eqnarray}
where $k_B T_h$ is the average energy deposited in the gas per recombination event.

\begin{figure}
\includegraphics[width=3.2in]{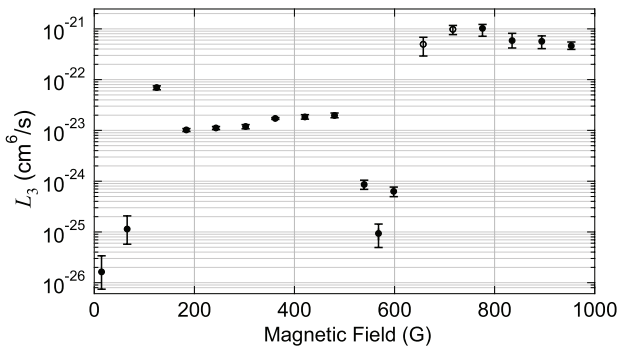}
\caption{\label{L3Graph} Three-body loss rate coefficient, $L_3$, at various fields.\vspace{-0.15in}}
\end{figure}

For a given parameter set ($N(0)$, $T(0)$, $L_1$, $L_3$ and $T_h$),
Eqns.~\ref{NumEvol} and ~\ref{TempEvol} can be
numerically integrated to give $N(t)$ and $T(t)$.
The best-fit parameters for a data set
at a given field are determined by minimizing $\chi^2$.
The best-fit $L_3$ values at various fields are displayed
in Fig.~\ref{L3Graph}.  For these fits, $T_h$ varied from $1.5 - 4\,\mu{\mathrm{K}}$ and $L_1$ varied from
$0.1 - 0.33\,{\mathrm{s}}^{-1}$.  The error bars in Fig.~\ref{L3Graph} indicate the 68\% confidence interval due to statistical uncertainty in all fit parameters.  The error bars do not include a
systematic uncertainty of $\pm70\%$ arising from our
uncertainty in atom number ($\pm 30\%$) and trap frequencies ($\pm 5\%$).

%
%

As shown in Fig.~\ref{L3Graph}, $L_3$ varies
by over four orders of magnitude.  In the range $600\,\mathrm{G} < B < 750\,\mathrm{G}$, the $L_3$ values ($\circ$) may be
overestimated since the measured decay includes loss events we observe in two-state mixtures at these field values.
Near 0 and 568~G, three-body recombination rates are relatively small and stable
three-state Fermi gases can be created.  For $L_3 = 10^{-25}\,\mathrm{cm}^{6}/\mathrm{s}$, a gas with a density
$n = 10^{12}\,\mathrm{cm}^{-3}$ per spin state has a
$1/e$-lifetime $\sim 10$ seconds.
Consistent  with the data shown in Fig.~\ref{FractionRemain}(a), a resonant peak in
$L_3$ is observed at $B\simeq 127\,\mathrm{G}$.  We
do not report a value for $L_3$ at the location of the 504~G resonant feature
(Fig.~\ref{FractionRemain}(a)) since a population imbalance develops there.
The highest $L_3$ value we report
($10^{-21}\,\mathrm{cm}^{6}/\mathrm{s}$) occurs at 775~G and is likely
 unitarity limited (for fields $>600$~G, $T \simeq 6\,\mu\mathrm{K}$ due to heating).  We
observe that $L_3$ decreases by a factor of $\simeq 2.5$ as the field is
increased from 775 to 953 G, the highest field we currently access.
This trend suggests that $L_3$ may decrease further in the high field limit
where $a_{12},\,a_{23},\,a_{13} \rightarrow -2140 a_0$.

In conclusion, we can create long-lived three-state DFGs of $^6$Li
atoms for fields near 568~G.  The minimal two-body loss we observe
makes this system promising for future studies of Efimov trimers.  Indeed, the loss resonances at
127 and 504~G may be occurring at field values where
trimer states cross the dissociation threshold.
The recombination rates at high fields need not preclude experiments
designed to study color superfluidity
and trion formation~\cite{Hofstetter07}.  For instance, using low density
gases ($n \sim 5\times10^{10}\,\mathrm{cm}^{-3}$) and
large period ($d \simeq 2\,\mu\mathrm{m}$) optical lattices, long lifetimes
($\gtrsim 0.1\,\mathrm{s}$) and strong interactions can simultaneously be achieved.

This material is based upon work supported by the AFOSR (Award FA9550-08-1-0069), the ARO (Award W911NF-06-1-0398) and
the NSF (PHY 07-01443).


\begin{thebibliography}{39}
\bibitem[{Joc()}]{JochimObservation}
\bibinfo{journal}{This resonance was first reported in Ref.[13]}.
\bibitem[{Tra()}]{TrapFrequencies}
\bibinfo{journal}{We determine trap frequencies by observing dipole
oscillations ($\omega_y$) and parametric excitation ($\omega_x$, $\omega_z$).}
\bibitem[{Tra()}]{FieldUncertainty}
\bibinfo{journal}{Fields are calibrated to $\pm 1{\mathrm{G}}$ by RF/optical spectroscopy.}
\bibitem[{Med()}]{Mediator}
\bibinfo{journal}{Three particles strongly interact even if one of the pairwise
scattering lengths is small since the third particle acts as a mediator for
the weakly interacting pair.}
\end{thebibliography}
\end{document}